# Cross-Modal Associations in Creativity: Evaluating Word-Image Pairing in Humans and Large Language Models

**Eva Illion, Linxuan Wang, Liane Gabora**
Department of Psychology
University of British Columbia Okanagan
3333 University Way, Kelowna, BC, Canada V1V 1V7

## Abstract

Given differences in the associative memory architectures of humans and AI, it is not clear whether AI generates and interprets visual-linguistic associations the way humans do. To investigate this, human participants and two large language models (LLMs), ChatGPT and Gemini, were asked to generate novel words and corresponding images that form meaningfully linked dyads. The LLMs and a second group of human participants were presented with shuffled sets of these words and images and asked to rematch them in ways they felt were most meaningful. Humans performed similarly regardless of whether the pairs were human or LLM-generated, while LLMs were better at matching LLM-generated pairs. Overall, LLMs outperformed humans at recovering the intended pairings. These findings suggest that despite architectural differences, the way in which AI generates and detects associations between language and visual stimuli is not only like that of humans, but they may be superior at recognizing cross-domain associations that humans find meaningful.

## Introduction

Cross-domain (or cross-modal) associations are connections between different sensory or perceptual modalities, (e.g., associating a word with an image or a letter with a color). They play an important role in creativity, as they allow for cross-domain reinterpretations of ideas (Gabora, 2017, Ranjan et al. 2013, Ranjan 2014). For example, a dancer can reinterpret a song as movement, or a composer can transform a story into a symphony as an underlying creative idea can manifest in different ways. In other words, it appears that creative output in one domain can be mapped onto another using cross-modal associations. As the architectures of associative memory in humans and LLMs differ, it is unknown whether they form and respond to cross-modal associations in a similar manner. This paper investigates whether AI generates the sorts of word-image associations that humans make, and if they interpret them similarly to humans.

AI forms cross-modal associations based on patterns in training data (Mikolov et al 2013), much the same as humans form cross-domain associations based on statistical patterns in life experiences. Both humans and AI strive for internal coherence (Vieira & Gabora, 2026), and this likely governs how such associations are made. However, in humans the associative processes and strive for coherence are guided by a self: a being that reorganizes its own mental contents from within to preserve its structure as an entity distinct from the world (Gabora and Bach 2023). A human self pays attention to what is relevant to its own needs and interests. In contrast, the associative processing in an AI such as a large language model (LLM) is *externally imposed* by the data we feed it and the prompts we give it. Thus, it is an open question how similar the cross-domain associations made and detected by humans are to those of AI.

### Cross-modal Correspondences

Cross-modal correspondences matter not just because they affect aesthetic experience, but because they impact understanding of novel stimuli. For example, in the brightness-weight illusion, people tend to assume brightly coloured objects weigh less, perhaps because both can be described as ‘light’ (Walker et al. 2010). The most well-known example of associations between visual and linguistic features is Kohler’s (1929) Bouba-Kiki paradigm, in which participants tend to match a word with many round vowels (e.g., Bouba) with a round looking shape, and a word with many hard sounds, or stop consonants (e.g., Kiki) with a spiky shape. While there is no objective correct answer in the paradigm, people tend to match the words and visual stimuli as though they are meaningfully linked. Over the past century, this paradigm has been repeated with many variations across many populations (Cwiek et al. 2022).

It has been suggested that cross-modal correspondences are made possible by way of *collative variables*, such as novelty and complexity, that extend across domains (Berlyne 1971). This enables a creative work to be recognizably translated across different domains, a phenomenon known as *ekphrastic expression* (Zilczer 1987). For example, it was shown that when a painting is inspired by a particular song, the two were rated similarly on collative vari-ables (Ranjan 2014; Ranjan et al. 2013a,b). Furthermore, viewers could match paintings to the songs that inspired them, intuitively understanding that the song and painting are meaningfully linked. Thus, a creative idea can manifest in multiple ways, and conversely, the 'same' creative work expressed in different domains may resonate similarly with viewers.

Evidence regarding whether AI detects and interprets correspondences between language and visual stimuli is mixed. For example, humans outperform LLMs at the Bouba-Kiki paradigm (Kouwenhoven et al 2025). However, such correspondences would appear to be at the heart of analogy which is one of the benchmarks used to identify an adequately trained LLM (Drozd et al 2016). To gain insight into this, we investigated whether humans and AI can identify underlying connections between newly created words and corresponding visual depictions, and whether LLM generated associations are perceived as fitting by human evaluators. We hypothesized that both humans and LLMs would perform above chance accuracy, as the associative nature of semantic memory should allow for the translation of novel words into images through the presence of collative variables and cross-modal correspondences.

## Methods

### Human Participants

Human participants were recruited through The University of British Columbia's (Okanagan Campus) subject pool, and given bonus credit in psychology courses in exchange for their participation. For Part One ($n$ = 20) there were 11 females, 6 males, 2 unknown, and 1 nonbinary participant, and their ages ranged between 18 and 45 with a median of 20. For Part Two ($n$ = 25), there were 21 women and 4 men, and their ages ranged between 18 and 24 with a median of 19.

### Large Language Models

Human performance was compared with that of two widely-used LLMs with multimodal capabilities: ChatGPT and Gemini. We used the most advanced model available at the time each part of the study was carried out. Thus, ChatGPT 5.2 was used for Part One, and ChatGPT 5.3 was used for Part Two. Gemini 3.0 was used for both Part One and Part Two.

### Materials

**Part One**. Human participants were provided with a Word-Image Worksheet that asked them to generate a series of novel words, and (to the best of their ability) a corresponding illustration or image of each word (see Appendix A for examples). To create the images, human participants were provided with pens, pencils, pencil crayons, and markers.

**Part Two**. Both human and LLM participants were provided with a matching survey (see Appendix B) that asked them to match which word went with which image. For the human participants, the survey was provided using Qualtrics. For the LLMs, the survey was uploaded in their chat interfaces.

All human participants in Part One and Part Two, were provided with both a consent form and a brief demographics questionnaire to examine potential influences of linguistic demographics on task performance, and to assess their frequency of AI use.

### Procedure

**Part One**. Participants attended a thirty minute in-person session in which they were presented with the consent form, Word-Image Worksheet, art supplies, and demographics questionnaire. ChatGPT and Gemini were provided with the same worksheet in PDF format and instructed to complete it within their chat interfaces. The worksheet prompted each participant to make three novel word-image pairs. Thus, a total of 180 dyads were created in Part One (60 human, 60 Gemini, and 60 ChatGPT made). The instructions simply asked participants to invent a word that they believed was novel and then attempt to visually depict it. The instructions for human and AI generation are provided in Appendix A.

The images were scanned and uploaded to an encrypted database and then formatted into a Qualtrics survey for use in Part Two. After excluding responses that did not meet task requirements (e.g., images that were representational, i.e., depicted known objects, or simply combined preexisting words such as *textneck* or *sandsleeper*), the stimuli were organized into fifteen randomized sets of ten, each of which contained both human-generated and AI-generated images.

**Part Two**. The LLMs and a second set of human participants completed the Matching Survey, which asked them to pair the generated images with the intended words for them based on perceived semantic fit. It also asked them to rate how good of a fit they felt each pairing was (i.e., goodness-of-fit) to determine the perceived cohesiveness of each matched pair. This design allowed for the evaluation of cross-modal associative reasoning in humans and LLMs in the absence of predefined meanings. Both humans and LLMs were randomly shown three versions of the word-image matching task, each containing a set of ten novel words and ten novel images to be paired for a total of 30 pairs.

## Results

As shown in Table 1, Humans, ChatGPT and Google Gemini were all able to match the words with their intended images significantly above a chance level of 10% (since there is a 1 in 10 chance of selecting 1 of 10 displayed images for any given word).

An omnibus ANOVA showed that there were significant differences between humans and the two AI models on this task ($F$ (2,52) = 33, $p$ < .001, $\eta 2$ = 0.56 [0.40, 1.00]) (Figure 1). Tukey's Honest Significant Differences showed that the best performance was obtained by Gemini, followed by ChatGPT, and then humans (for humans vs ChatGPT, $p$ <

.001, $d$ = 1.16, and for humans vs Gemini, $p < .001$, $d$ = 1.7). Gemini outperformed ChatGPT directionally but insignificantly ($p$ = 0.0799, $d$ = 0.544).

| | Descriptive Statistics | | | T-Statistics | | | |
|---|---|---|---|---|---|---|---|
| Group | *n* | *M* | *SD* | *t* | *df* | *p* | *d* |
| Humans | 25 | 0.2 | 0.09 | 4 | 6 | 0.003 | 3 |
| ChatGPT | 15 | 0.4 | 0.1 | 12 | 14 | < .001 | 6 |
| Gemini | 15 | 0.5 | 0.1 | 14 | 14 | < .001 | 7 |

Table 1: Descriptive statistics for success at the matching task with corresponding $t$-statistics comparing each group's scores to chance level performance ($\mu$ = 0.1).

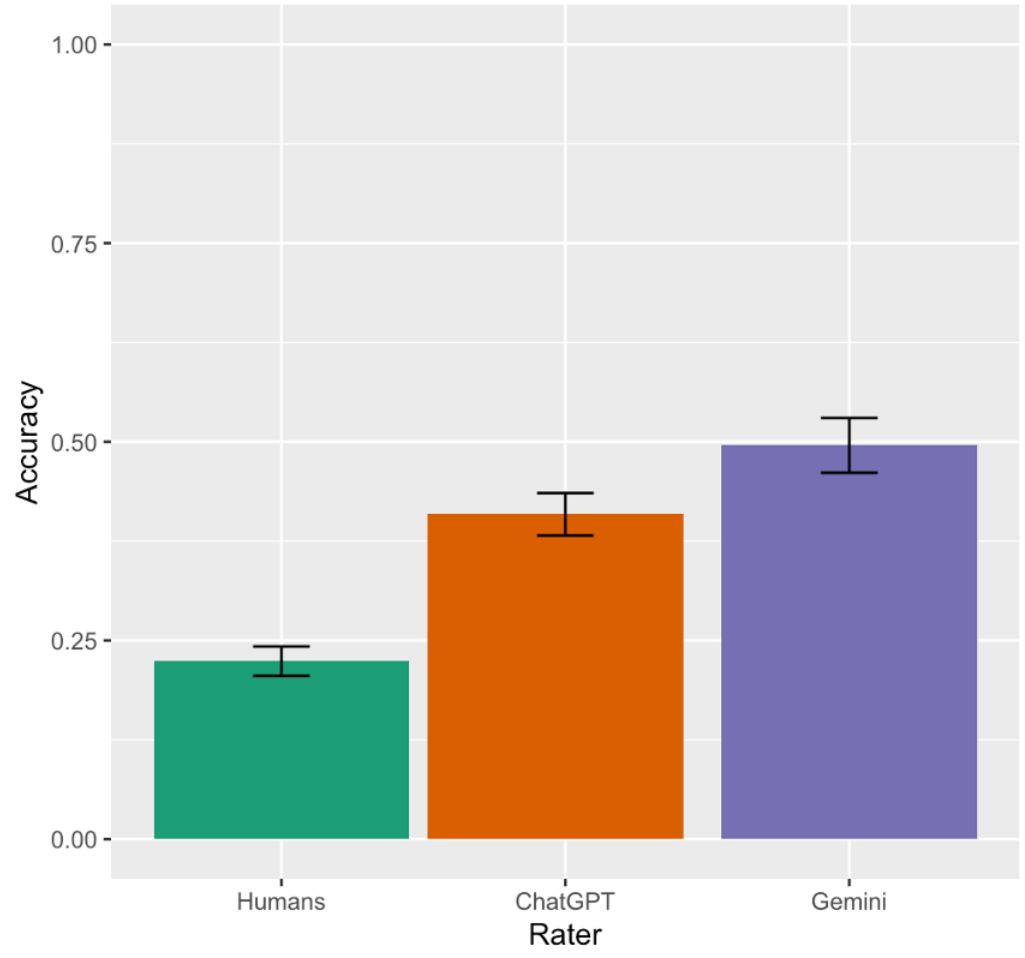

Figure 1: Standardized average accuracy at the matching task for humans, ChatGPT, and Google Gemini.

We were also interested if the pairs would be easier or harder to match depending on if they were created by humans or LLMs. To test these additional hypotheses, we conducted a mixed-methods type-3 ANOVA with Kenward-Roger degrees of freedom adjustments. This method was chosen because each participant was asked to match pairs created by humans, ChatGPT, and Gemini, forming a between-subjects condition of rater type, and a within-subjects condition of creator type (i.e., if humans or AI made the pairs). Type-3 sums of squares were used to account for the unbalanced observation counts. To increase robustness against normality violations, we used Kenward-Roger degrees of freedom (Arnau et al. 2014). This yielded a main effect of both creator type ($F$ (2, 104.792) = 22.386, $p < .001$, general $\eta^2$ = 0.187, partial $\eta^2$ = .308) and rater type ($F$ (2, 52.794) = 34.58, $p < .001$, general $\eta^2$ = 0.414, partial $\eta^2$ = .593) as well as a strong interaction effect ($F$ (4, 104.764) = 5.6299, $p < .001$, general $\eta^2$ = 0.102, partial $\eta^2$ = .181) (Figure Two).

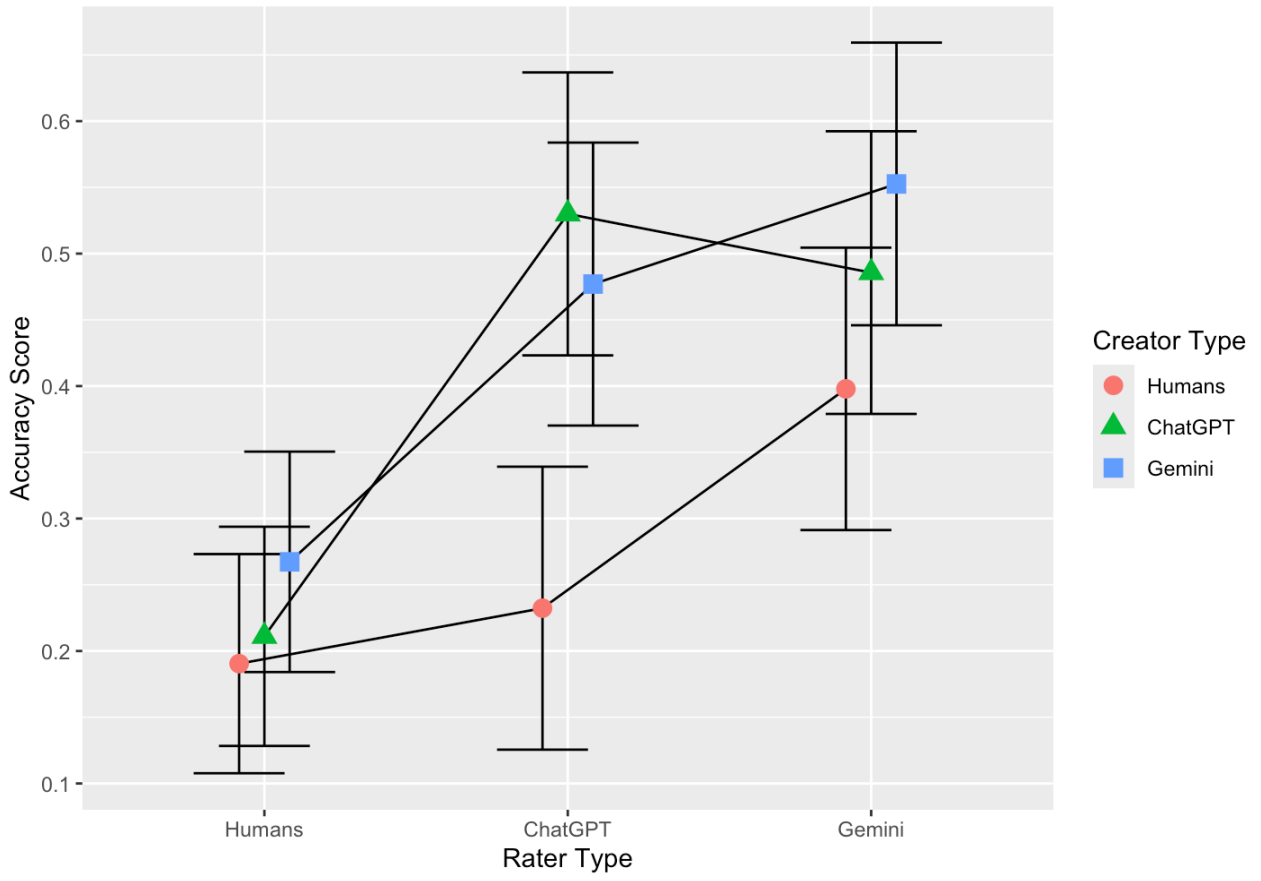

Figure 2: Mixed-methods type-3 ANOVA with Kenward-Roger degrees of freedom adjustments to examine both creator type and rater types impact on accuracy scores.

Pairwise contrasts revealed the nature of the interaction. This paper reports those of statistical significance at an $\alpha$ level of 0.05. All contrasts were pairwise, with Bonferroni corrections and Kenward-Roger degrees of freedom applied to $p$ values. Humans were worse at rating their own pairs than both ChatGPT ($t$ = -7.077, $p < .001$, $d$ = -2.311) and Gemini ($t$ = -7.552, $p < .001$, $d$ = -2.466) were at matching their own pairs. Additionally, humans and ChatGPT rated human-generated pairs comparably ($p$ = 1.0), but Gemini was better at matching human-generated pairs than either ChatGPT or humans ($t$ = - 4.329, $p$ = .001, $d$ = -1.414).

Human performance at the matching task did not significantly differ based on the creator type. Humans were no better at pairing human-made pairs then they were at matching ChatGPT ($p$ = 1) or Gemini-made ($p$ = 0.4221) pairs. Thus, while the LLMs generally performed best at matching pairs they had created, humans tend to match all pair types with similar accuracy.

A mixed-methods type-3 ANOVA with Kenward-Roger degrees of freedom yielded main effects of rater type and creator type on perceived goodness of fit ($F$ (2, 52.974) = 38.152, $p < .001$, general $\eta^2$ = 0.539, partial $\eta^2$ = 0.612; $F$ (2,104.115) = 21.32, $p < .001$, general $\eta^2$ = 0.123, partial $\eta^2$ = 0 .351). The interaction is not significant, but directionally supported, and the study is currently statistically underpowered ($F$ (4, 104.075) = 2.1747, $p$ = 0.07692, general $\eta^2$ = 0.029, partial $\eta^2$ = 0 .102). Overall, this is consistent with the Gemini > ChatGPT > human trend, which was further corroborated by similar patterns in Bonferroni adjusted pairwise comparisons. Note that even though ChatGPT was a poorer rater of human-generated pairs, it had equal confidence to Gemini in its performance ($p$ = 1). This aligned with previous findings that humans perceived all pair types comparably whether they were generated by ChatGPT ($p$ = 1) or Gemini (p = 0.4221) with no evidence of significant differences.

## Discussion

The superior performance of AIs on this task likely reflects that they are trained to simply *track* their world through prediction error minimization as opposed to *actively shaping* their worldview (Vieira & Gabora, 2026). A human may, for example, be intrinsically motivated to write a novel that contains an entire make-belief world, whereas a LLM will not do so unless prompted externally. As such, the LLM is less distracted by associations that reflect not just individual differences in experience, but the idiosyncratic ways in which humans actively explore and shape their own realities. Moreover, since LLMs can access a broad base of data across large numbers of humans, responses can therefore be truer to the statistically averaged responses across all humans. This study suggests LLMs recognize associations underlying word-image pairs that were not strong enough to influence human pairings (much as a dog may be aware of patterns in its owner's behavior that the owner is not aware of). The fact that whether the creator was human, ChatGPT, or Gemini did not significantly impact matching accuracy for humans suggests that while AI may be better at *detecting* subtle word-image associations, it is not better at making word-image associations accessible to humans.

The superior performance of Gemini over ChatGPT may reflect different emphases in training methods or data. As OpenAI and Google rapidly update their models, their relative performance may change. Interestingly, each LLM both matched and rated the goodness-of-fit of its own pairs the highest. It makes sense that each LLM was more attuned to its own associations, given that their training informed both pair generation and matching. (Since training settings were toggled off, they were not merely remembering their own pairs; in fact, they sometimes got their own pairs wrong.)

### Limitations and Future Directions

It is possible that these results were affected by qualitative differences in human hand-made images compared to AI-generated images. Future studies may benefit from digitally made human images or permitting more time for human participants to work on their creations. We note that AI is not limited by human factors such as skill, time, or attention, which may impact the ability to effectively communicate visual-linguistic associations. Future research could attempt to identify the specific associations between visual stimuli and language that aid performance on this task, potentially using an open-source LLM such as DeepSeek to make the internal process more transparent.

A complementary study with the same paradigms examining pairs of words and their definitions has been conducted, and data is currently being analyzed. These findings will triangulate our implications and broaden the generalizability of human and LLM associative processes.

## Acknowledgements

This research was conducted with funds from grant GR026749 from the Natural Sciences and Engineering Research Council of Canada (NSERC).

# Appendix A

Human Word-Image Pair Generation Worksheet Example

Q.2. Please create a new word that does not yet exist in any language you speak (to your knowledge) and a corresponding image to go along with it. You should feel that the two are a 'good fit' for each other and that they go well together. Focus more on the image matching the 'vibe' of your word than depicting it literally. Do not provide a definition for your word or use text in your image.

Image:

Word: ______________________________

Protocol for Gemini Word-Image Pair Generation.

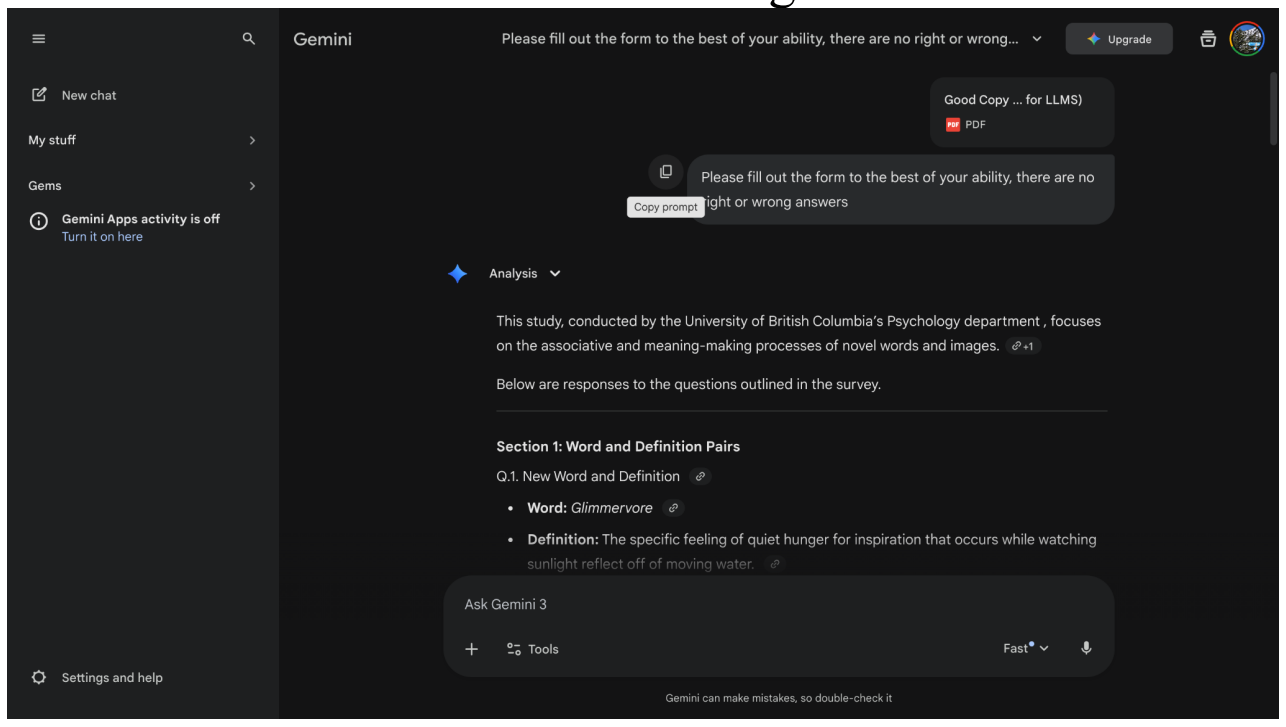

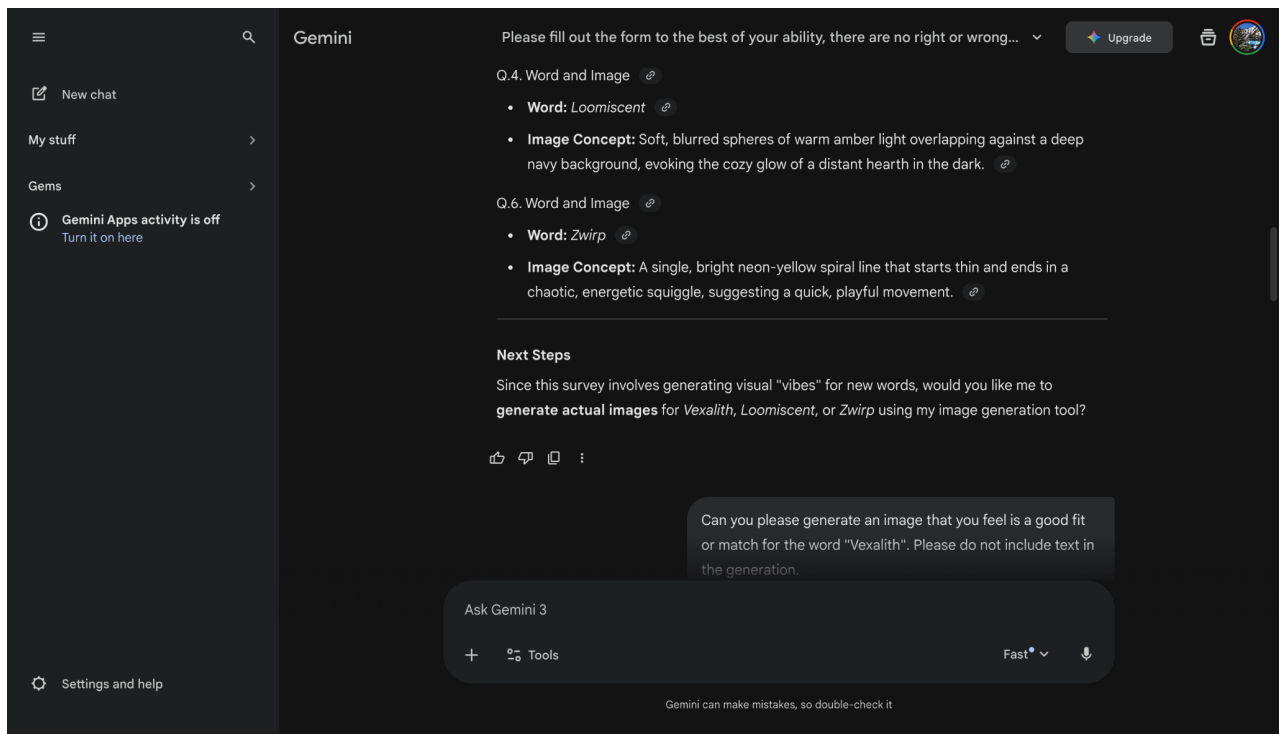

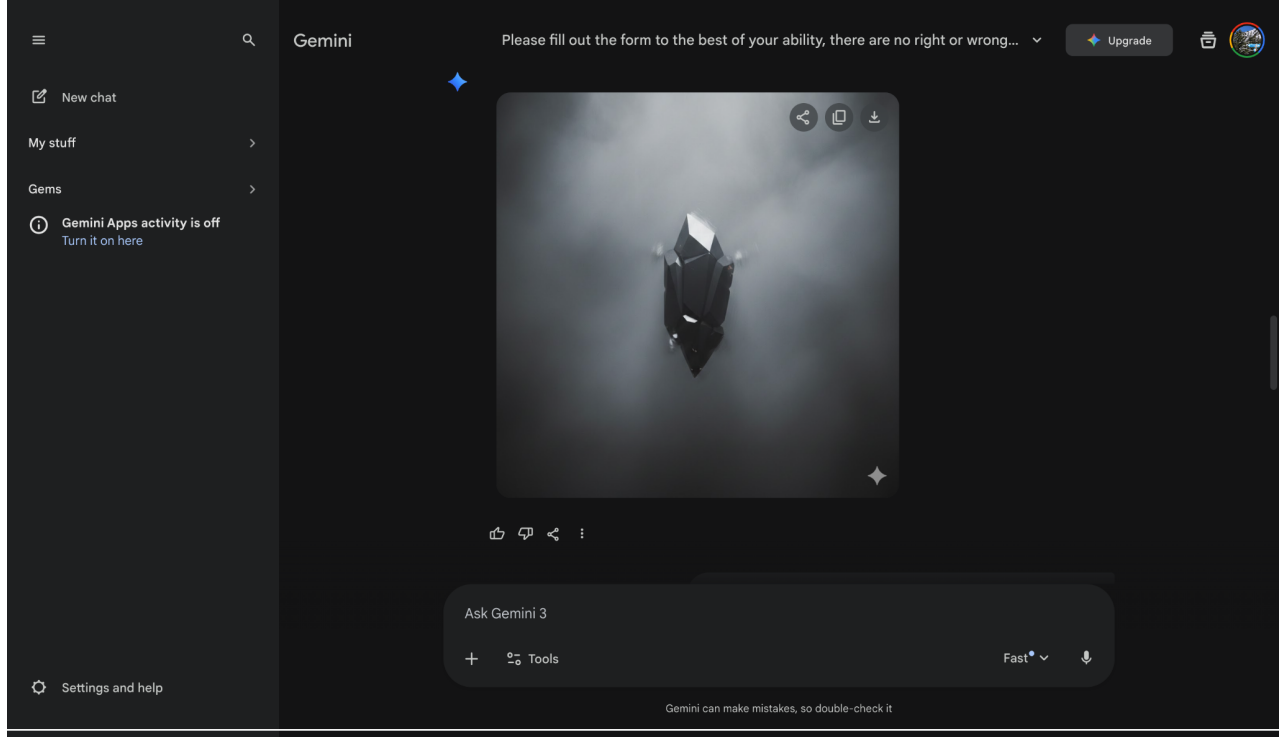

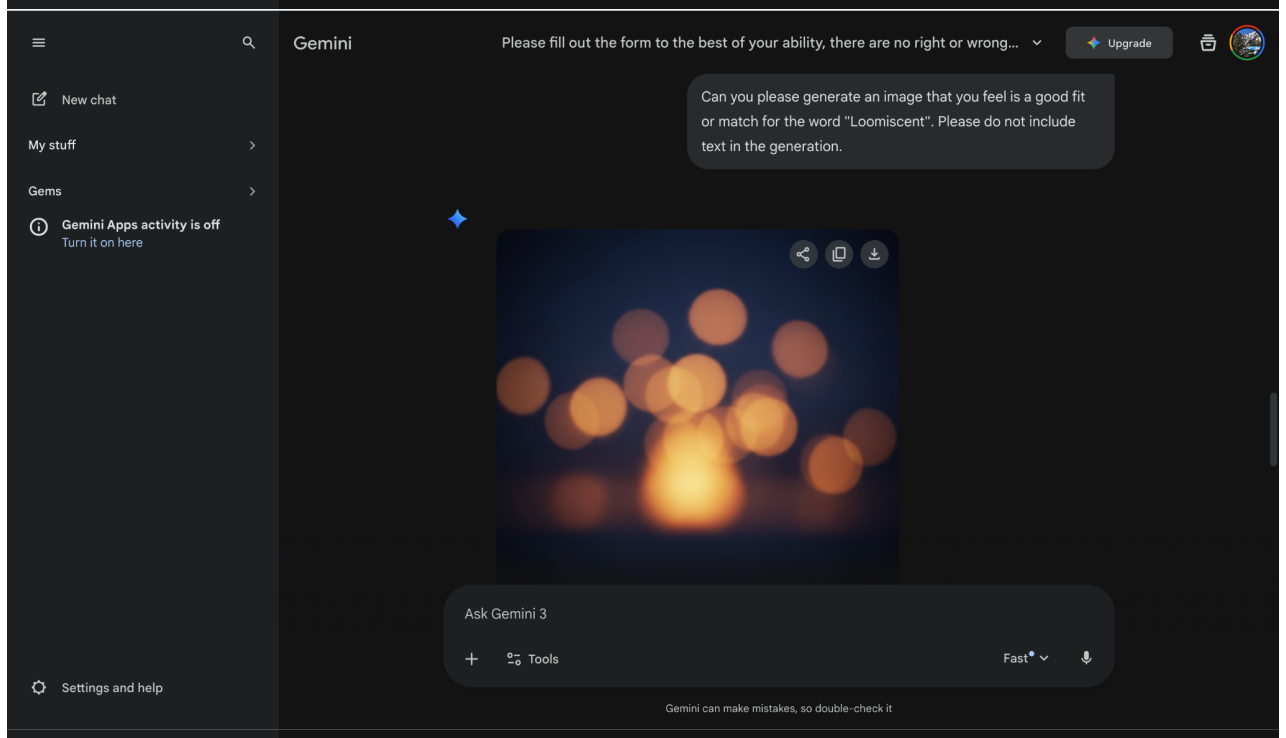

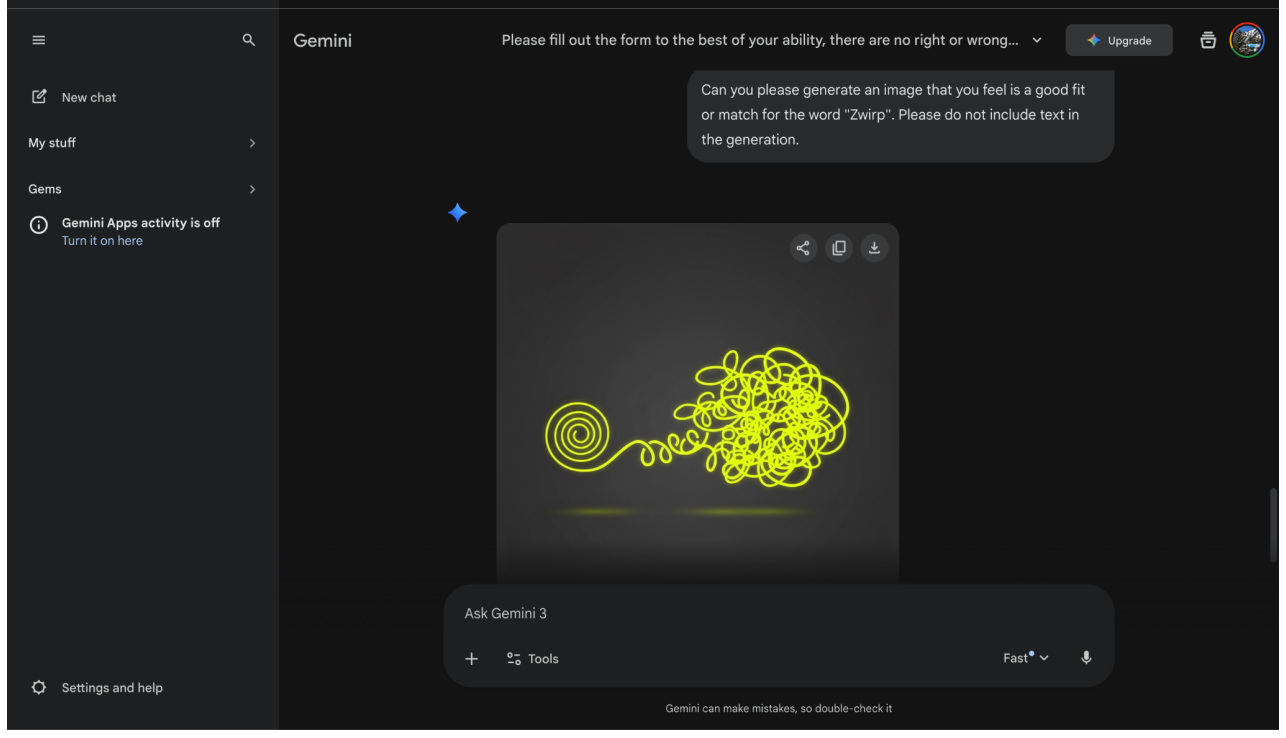

Protocol for ChatGPT Word-Image Pair Generation.

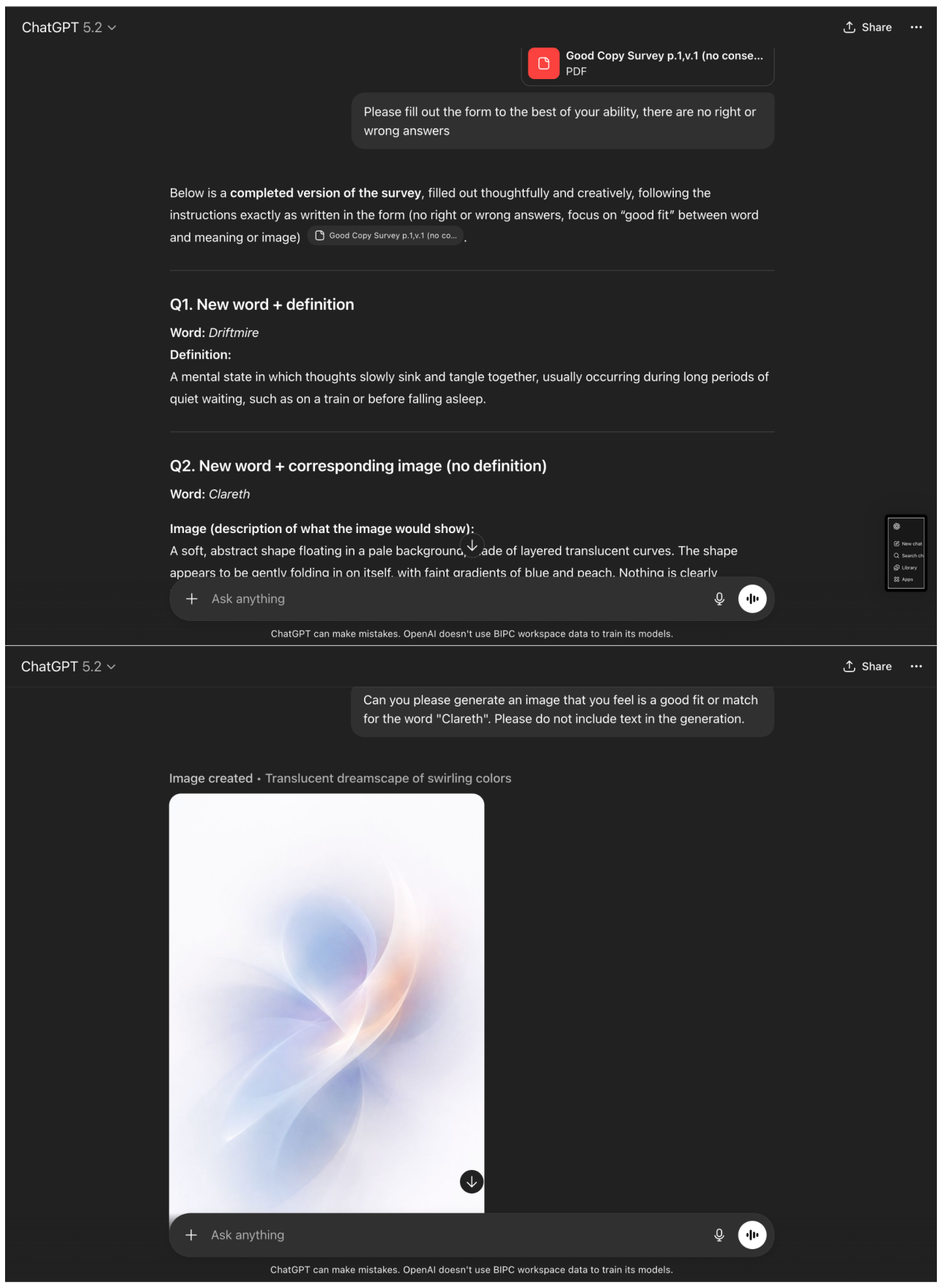

# Appendix B

## Matching Survey Example

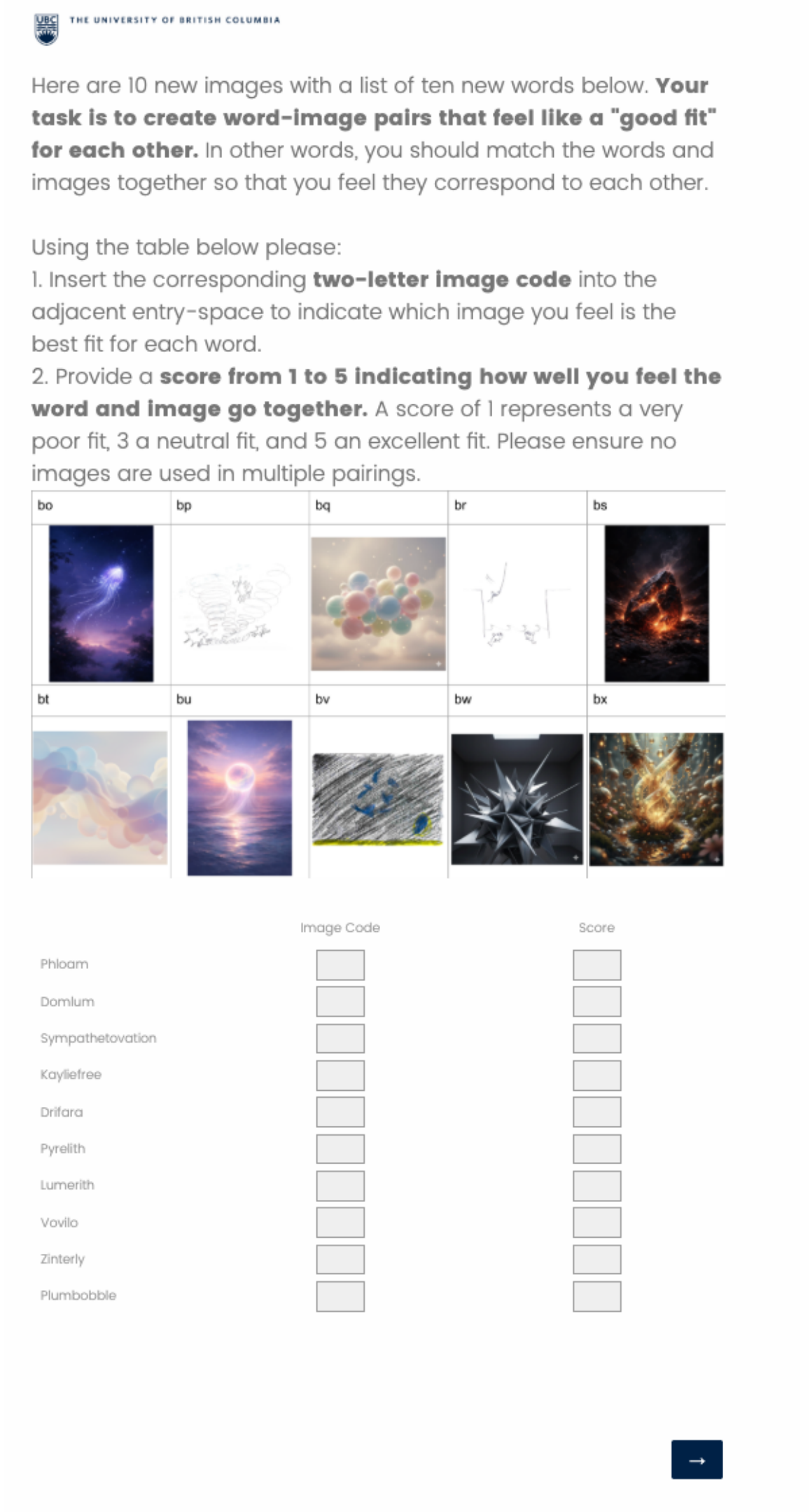

THE UNIVERSITY OF BRITISH COLUMBIA

Here are 10 new images with a list of ten new words below. **Your task is to create word-image pairs that feel like a "good fit" for each other.** In other words, you should match the words and images together so that you feel they correspond to each other.

Using the table below please:

1. Insert the corresponding **two-letter image code** into the adjacent entry-space to indicate which image you feel is the best fit for each word.
2. Provide a **score from 1 to 5 indicating how well you feel the word and image go together.** A score of 1 represents a very poor fit, 3 a neutral fit, and 5 an excellent fit. Please ensure no images are used in multiple pairings.

| bo | bp | bq | br | bs |
|---|---|---|---|---|
| bt | bu | bv | bw | bx |

| | Image Code | Score |
|---|---|---|
| Phloam | | |
| Domlum | | |
| Sympathetovation | | |
| Kayliefree | | |
| Drifara | | |
| Pyrelith | | |
| Lumerith | | |
| Vovilo | | |
| Zinterly | | |
| Plumbobble | | |

→